\documentclass[12pt, one column]{IEEEtran}
\usepackage{amsmath, amsthm, amsfonts}
\usepackage[english]{babel}
\usepackage{graphicx}
\usepackage{tabularx}
\usepackage{multirow}
\usepackage{enumerate}
\usepackage{setspace}
\usepackage[ruled,vlined]{algorithm2e}
\begin{document}

\title{High Speed VLSI Architecture for 3-D Discrete Wavelet Transform}

\author{{B.K.N.Srinivasarao and Indrajit Chakrabarti\\
Department of Electronics and Electrical Communication Engineering\\
Indian Institute of Technology, Kharagpur, INDIA\\
E.Mail : srinu.bkn@iitkgp.ac.in,
indrajit@ece.iitkgp.ernet.in}}
\maketitle
\thispagestyle{empty}
\doublespacing

\begin{abstract}

This paper presents a memory efficient, high throughput parallel lifting based running three dimensional discrete wavelet transform (3-D DWT) architecture. 3-D DWT is constructed by combining the two spatial and four temporal processors.  Spatial processor (SP) apply the two dimensional DWT on a frame, using lifting based 9/7 filter bank through the row rocessor (RP) in row direction and then apply in the colum direction through column processor (CP). To reduce the temporal memory and the latency, the temporal processor (TP) has been designed with lifting based 1-D Haar wavelet filter. The proposed architecture replaced the  multiplications by pipeline shift-add operations to reduce the CPD. Two spatial processors works simultaneously on two adjacent frames and provide 2-D DWT coefficients as inputs to the temporal processors. TPs apply the one dimensional DWT in temporal direction and provide eight 3-D DWT coefficients per clock (throughput). Higher throughput reduces the computing cycles per frame and enable the lower power consumption. Implementation results shows that the proposed architecture has the advantage in reduced memory, low power consumption, low latency, and high throughput over the existing designs. The RTL of the proposed architecture is described using verilog and synthesized using 90-nm technology CMOS standard cell library and results show that it consumes 43.42 mW power and  occupies an area equivalent to 231.45 K equivalent gate at frequency of 200 MHz. The proposed architecture has also been synthesised for the Xilinx zynq 7020 series field programmable gate array (FPGA).
\end{abstract}

\begin{IEEEkeywords}
Index Terms : discrete wavelet transform, 3-D DWT, lifting based DWT, VLSI Architecture, flipping structure, strip-based scanning.
\end{IEEEkeywords}

\section{Introduction}
\par  Video compression is a major requirement in many of the recent applications like medical imaging, studio applications and broadcasting applications. Compression ratio of the encoder completely depends on the underlying compression algorithms. The goal of compression techniques is to reduce the immense amount of visual information to a manageable size so that it can be efficiently stored, transmitted, and displayed. 3-D DWT based compressing system enables the compression in spatial as well as temporal direction which is more suitable for video compression. Moreover, wavelet based compression provide the scalability with the levels of decomposition. Due to continuous increase in size of the video frames (HD to UHD), video processing through software coding tools is more complex. Dedicated hardware only can give higher performance for high resolution video processing. In this scenario there is a strong requirement to implement a VLSI architecture for efficient 3-D DWT processor, which consumes less power, area efficient, memory efficient and should operate with a higher frequency to use in real-time applications. \\

\par From the last two decades, several hardware designs have been noted for implementation of 2-D DWT and 3-D DWT for different applications. Majority of the designs are developed based on three categories, viz. (i) convolution based (ii) lifting-based and (iii) B-Spline based.  Most of the existing architectures are facing the difficulty with larger memory requirement, lower throughput, and complex control circuit. In general the circuit complexity is denoted by two major components viz, arithmetic and Memory component. Arithmetic component includes adders and multipliers, whereas memory component consists of temporal memory and transpose memory. Complexity of the arithmetic components is fully depends on the DWT filter length. In contrast size of the memory component  is depends on dimensions of the image. As image resolutions are continuously increasing (HD to UHD), image dimensions are very high compared to filter length of the DWT, as a result complexity of the memory component occupied major share in the overall complexity of DWT architecture.\\

\par Convolution based implementations \cite{3D_conv_dai}-\cite{3D_conv_mohanty} provides the outputs within less time but require high amount of arithmetic resources, memory intensive and occupy larger area to implement. Lifting based a implementations requires less memory, less arithmetic complex and possibility to implement in parallel. However it require long critical path, recently huge number of contributions are noted to reduce the critical path in lifting based implementations. For a general lifting based structure \cite{dwt_1} provides critical path of $4T_{m} + 8T_{a}$, by introducing 4 stage pipeline it cut down to $T_{m} + 2T_{a}$. In \cite{3D_flip} Huang et al., introduced a flipping structure it further reduced the critical path to $T_{m} + T_{a}$. Though, it reduced the critical path delay in lifting based implementation, it requires to improve the memory efficiency. Majority of the designs implement the 2-D DWT,  first by applying 1-D DWT in row-wise and then apply 1-D DWT in column wise. It require huge amount of memory to store these intermediate coefficients. To reduce this memory requirements, several DWT architecture have been proposed by using line based scanning methods \cite{3D_huang}-\cite{3D_xiong}. Huang et al., \cite{3D_huang}-\cite{3D_huang2} give brief details of B-Spline based 2-D IDWT implementation and discussed the memory requirements for different scan techniques and also proposed a efficient overlapped strip-based scanning to reduce the internal memory size. Several parallel architectures were proposed for lifting-based 2-D DWT \cite{3D_huang2}-\cite{3D_yusong2}. Y. Hu et al. \cite{3D_yusong2}, proposed a modified strip based scanning and parallel architecture for 2-D DWT is the best memory-efficient design among the existing 2-D DWT architectures, it requires only 3N + 24P of on chip memory for a N$\times$N image with $ P $ parallel processing units (PU). Several lifting based 3-D DWT architectures are noted in the literature \cite{3D_zheng}-\cite{3D_darji} to reduce the critical path of the 1-D DWT architecture and to decrease the memory requirement of the 3-D architecture. Among the best existing designs of 3-D DWT, Darji et al. \cite{3D_darji} produced best results by reducing the memory requirements and gives the throughput of 4 results/cycle. Still it requires the large on-chip memory ($4N^{2}+10N$). 

In this paper, we propose a new parallel and memory efficient lifting based 3-D DWT architecture, requires only $2*(3N+60P)+ 48$ words of on-chip memory and produce 8 results/cycle. The proposed 3-D DWT architecture is built with two spatial 2-D DWT (CDF 9/7) processors and four temporal 1-D DWT (Haar) processors. Proposed architecture for 3-D DWT replaced the multiplication operations by shift and add, it reduce the CPD from $ T_{m}+T_{a} $ to $ 4T_{a} $. Further reduction of CPD to $T_{a}$ is done by introducing pipeline in the processing elements. To eliminate the temporal memory and to reduce the latency, Haar wavelet is incorporated in temporal processor. The resultant architecture has reduce the latency, on chip memory and to increase the speed of operation compared to existing 3-D DWT designs.  The following sections provide the architectural details of proposed 3-D DWT through spatial and temporal processors.  

\par Organization of the paper as follows. Theoretical background for DWT is given in section II. Detailed description of the proposed architecture for 3-D DWT  is provided in section III.  Implementation results and performance comparison is given in Section IV. Finally, concluding remarks are given in Section V.\\
\section{Theoretical background}
Lifting based wavelet transform designed by using a series of matrix decomposition specified by the Daubechies and Sweledens in \cite{dwt_1}. By applying the  flipping \cite{3D_flip} to the lifting scheme, the multipliers in the longest delay path are eliminated, resulting in a shorter critical path. The original data on which DWT is applied is denoted by $X[n]$, and the 1-D DWT outputs are the detail coefficients $H[n]$ and approximation coefficients $L[n]$.  For the Image (2-D) above process is performed in rows and columns as well.  Eqns.(1)-(6) are the design equations for flipping based lifting (9/7) 1-D DWT \cite{3D_flip2} and the same equations are used to implement the proposed row processor (1-D DWT) and column processor (1-D DWT). 
\begin{figure}[H]
\label{lift_2d_eq}
\begin{align}
{H_1}[n] &\leftarrow a'*X[2n-1]+\{X[2n]+X[2n-2]\}  \ldots P1\\
{L_1}[n]&\leftarrow b'*X[2n]+\{{H_1}[n]+{H_1}[n-1]\}  \ldots U1\\
{H_2}[n] &\leftarrow c'*{H_1}[n]+\{{L_1}[n]+{L_1}[n-1]\}\ldots P2\\
{L_2}[n] &\leftarrow d'*{L_1}[n]+\{{H_2}[n]+{H_2}[n-1]\}\ldots U2\\
H[n] &\leftarrow K0* \{{H_2}[n]\}\\
L[n] &\leftarrow K1* \{{L_2}[n]\}
\end{align}
\end{figure}
Where $ a'=1/\alpha $, $ b'=1/\alpha\beta $, $ c'=1/\beta\gamma $, $ d'=1/\gamma\delta $, $ K0= \alpha\beta\gamma/\zeta$, and $ K1= \alpha\beta\gamma\delta\zeta$  \cite{dwt_1}. The lifting step coefficients $ \alpha$, $ \beta$, $\gamma $, $ \delta $ and scaling coefficient $\zeta $ are constants and its values  $ \alpha = -1.586134342$, $ \beta =-0.052980118$, $\gamma =0.8829110762$, and $ \delta =0.4435068522$, and  $\zeta  = 1.149604398.$

Lifting based wavelets are always memory efficient and easy to implement in hardware. The lifting scheme consists of three steps to decompose the samples, namely, splitting, predicting (eqn. (1) and (3)), and updating (eqn. (2) and (4)). 

Haar wavelet transform is orthogonal and simple to construct and provide fast output. By considering the advantages of the Haar wavelets, the proposed architecture uses the Haar wavelet to perform the 1-D DWT in temporal direction (between two adjacent frames). Sweldens \textit{et al.} \cite{daub_Haar} developed a lifting based Haar wavelet. The equations of the lifting scheme for the Haar wavelet transform is as shown in eqn.(\ref{eq2})
\begin{equation}
\label{eq2}
\left[ \begin{array}{l}
L\\
H
\end{array} \right] = \left( {\begin{array}{*{20}{c}}
{\sqrt 2 }&0\\
0&{\frac{1}{{\sqrt 2 }}}
\end{array}} \right)\left( {\begin{array}{*{20}{c}}
1&{S(z)}\\
0&1
\end{array}} \right)\left( {\begin{array}{*{20}{c}}
1&0\\
{ - P(z)}&1
\end{array}} \right)\left( \begin{array}{l}
{X_0}(z)\\
{X_1}(z)
\end{array} \right)
\end{equation}
\begin{equation}
\label{eq3}
\begin{array}{l}
L = {\textstyle{1 \over {\sqrt 2 }}}({X_0} + {X_1})\\
H = {\textstyle{1 \over {\sqrt 2 }}}({X_1} - {X_0})
\end{array}
\end{equation}
Eqn.(\ref{eq3}) is extracted by substituting Predict value $P(z)$ as 1 and Update step $ S(z) $ value as 1/2 in eqn.(\ref{eq2}), which is used to develop the temporal processor to apply 1-D DWT in temporal direction ($ 3^{rd} $ dimension). Where L and H are the low and High frequency coefficients respectively.

\section{Proposed architecture for 3-D DWT}

\begin{figure} 
\centering
\includegraphics [height=80mm,width=70mm]{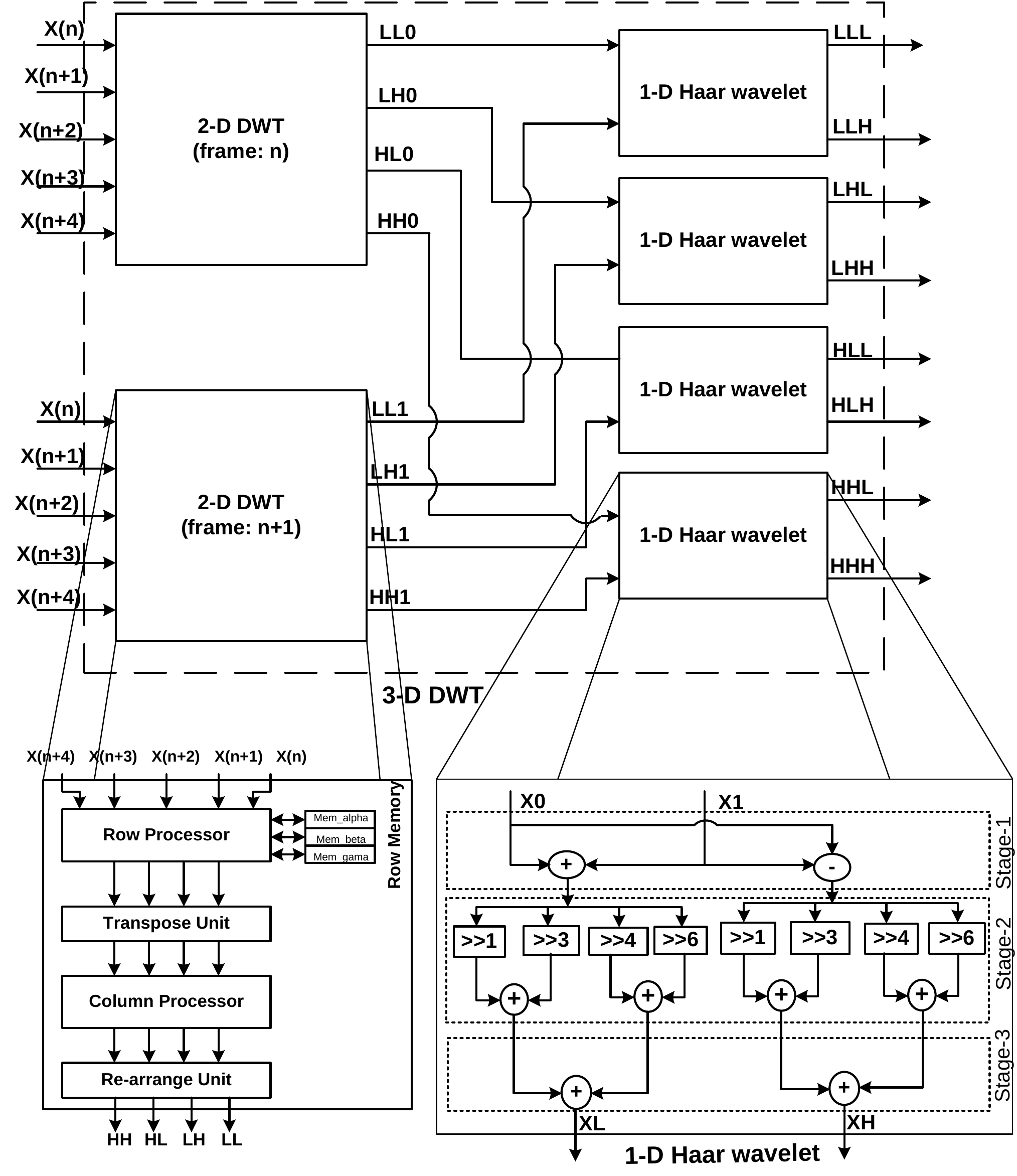}
\caption{Block diagram for 3-D DWT}
\label{blockdia_1}
\end{figure}

The proposed architecture for 3-D DWT comprising of two parallel spatial processors (2-D DWT) and four  temporal processors (1-D DWT), is depicted in Fig. \ref{blockdia_1}.  After applying 2-D DWT on two consecutive frames, each spatial processor (SP) produces 4 sub-bands, viz. LL, HL, LH and HH and are fed to the inputs of four temporal processors (TPs) to perform the temporal transform. Output of these TPs is a low frequency frame (L-frame) and a high frequency frame (H-frame). Architectural details of the spatial processor and temporal processors are discussed in the following sections.

\subsection{Architecture for Spatial Processor}
\begin{figure} 
\centering
\includegraphics [height=100mm,width=120mm]{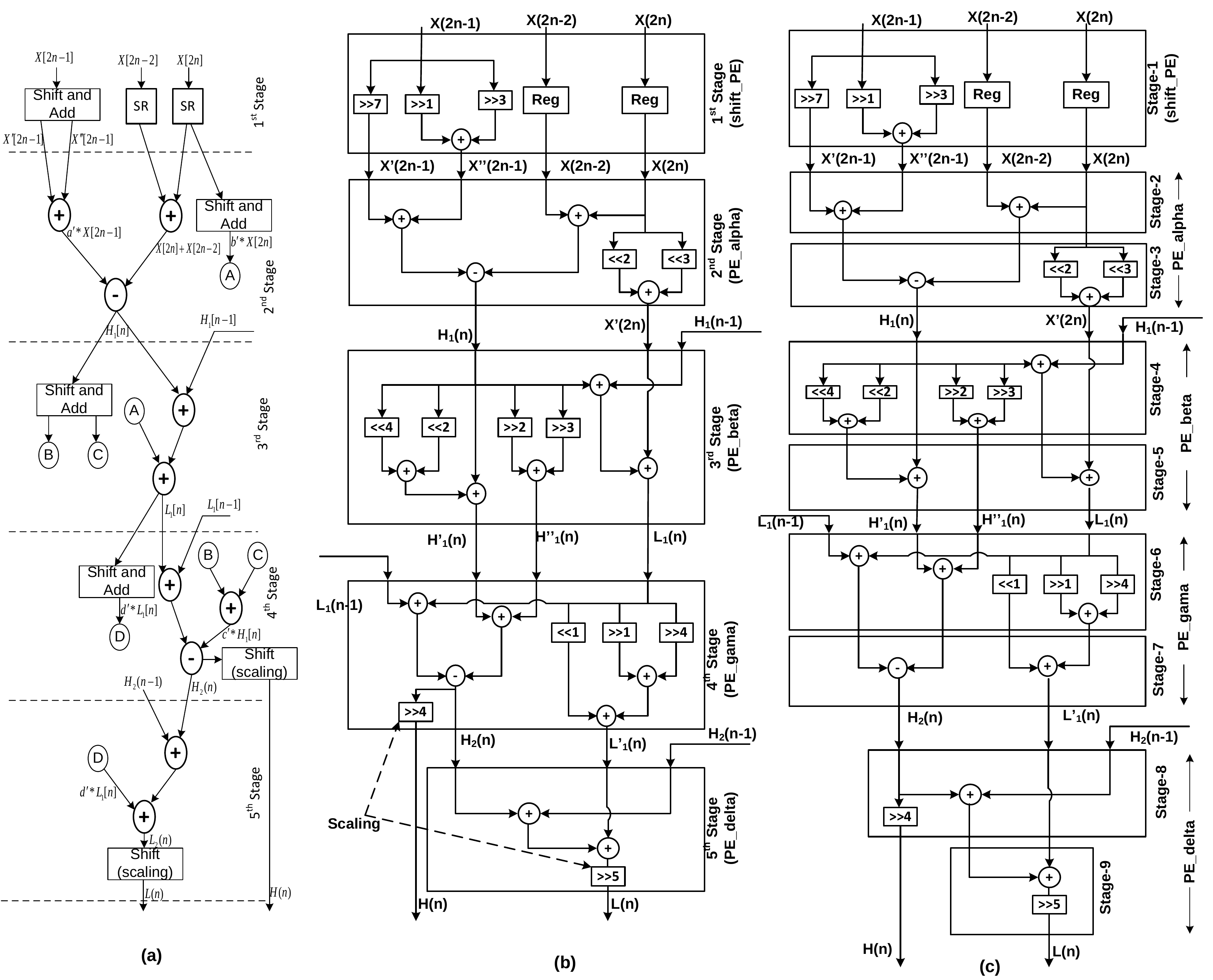}
\caption{(a) Data Flow Graph of Processing Unit (b) Processing Unit with five pipeline stages (c) Processing Unit with nine pipeline stages}
\label{3d_2}
\end{figure}

In this section, we propose a new parallel and memory efficient lifting based 2-D DWT architecture denoted by spatial processor (SP) and it consists of row and column processors. The proposed SP is a revised version of the architecture developed by the Y. Hu et al.\cite{3D_yusong2}. The proposed architecture utilizes the strip based scanning \cite{3D_yusong2} to enable the trade-off between external memory and internal memory. To reduce the critical path in each stage flipping model \cite{3D_flip}-\cite{3D_flip2} is used to develop the processing element (PE). Each PE has been developed with shift and add techniques in place of multiplier. Lifting based (9/7) 1-D DWT process has been performed by the processing unit (PU) in the proposed architecture. As shown in Fig.~\ref{3d_2}, the proposed PU is designed with five PEs, and each PE (except first PE (shift$\_$PE)) has been constructed with two pipeline stages for further reduction of CPD. This modified PU, reduces the CPD to $ T_{a} $ (adder delay). Fig.~\ref{blockdia_1} shows that the number of inputs to the spatial processor is equal to 2P+1, which is also equal to the width of the strip. Where P is the number of parallel processing units (PUs) in the row processor as well as column processor. We have designed the proposed architecture with two parallel processing units (P = 2). The same structure can be extended to P = 4, 8, 16 or 32 depending on external bandwidth. Whenever row processor produces the intermediate results, immediately column processor start to process on those intermediate results. Row processor takes 9 clocks to produce the temporary results then after column processor takes 9 more clocks to to give the 2-D DWT output; finally, temporal processor takes 3 more clocks after 2-D DWT results are available to produce 3-D DWT output. As a summary, proposed 2-D DWT and 3-D DWT  architectures have constant latency of 18 and 21 clock cycles respectively, regardless of image size N and number of parallel PUs (P). Details of the row processor and column processor are given in the following sub-sections.

\begin{figure} 
\centering
\includegraphics [height=120mm,width=130mm]{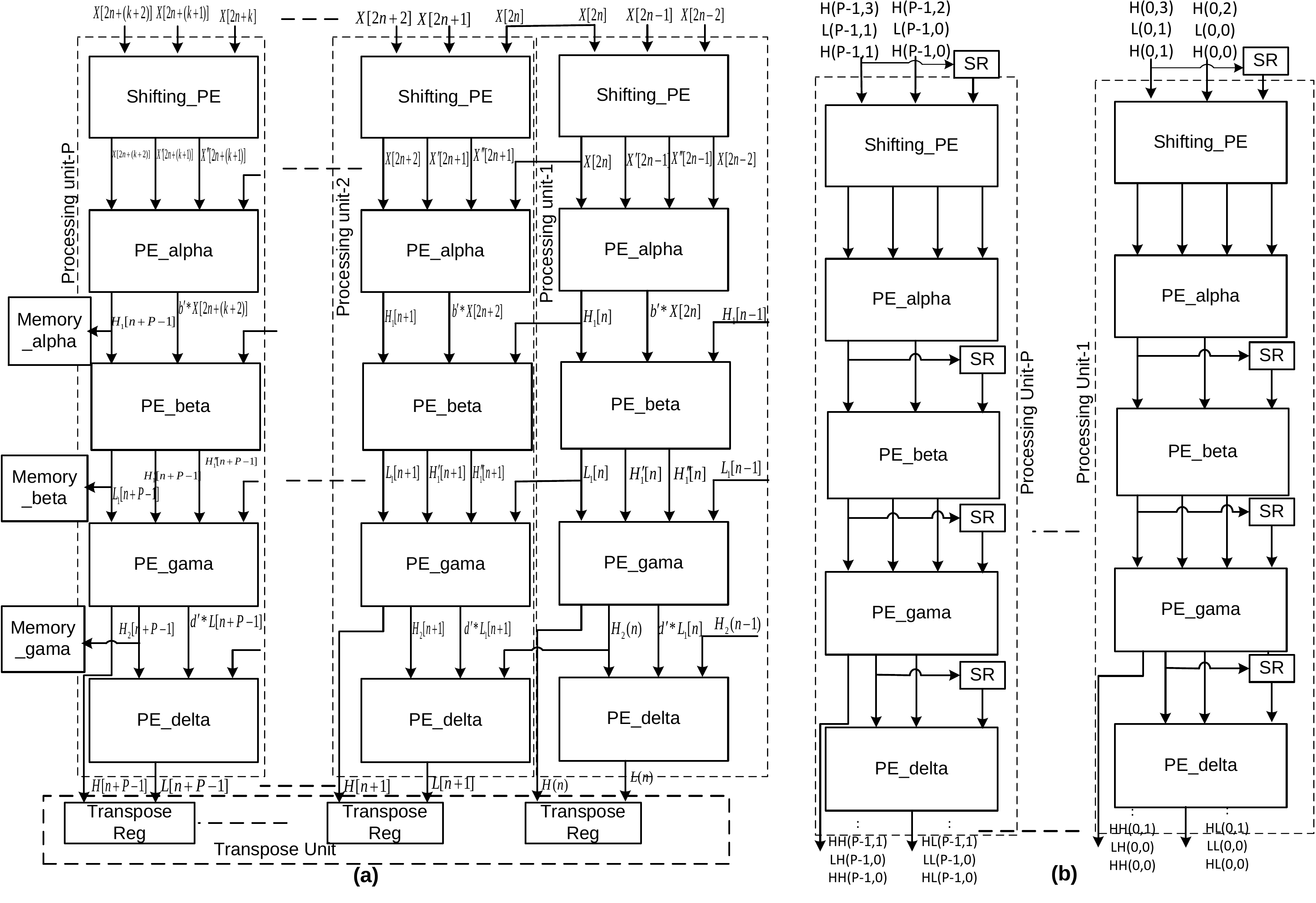}
\caption{(a)Row Processor  (b) Column Processor}
\label{3d_1}
\end{figure}

\begin{figure} 
\centering
\includegraphics [height=70mm,width=50mm]{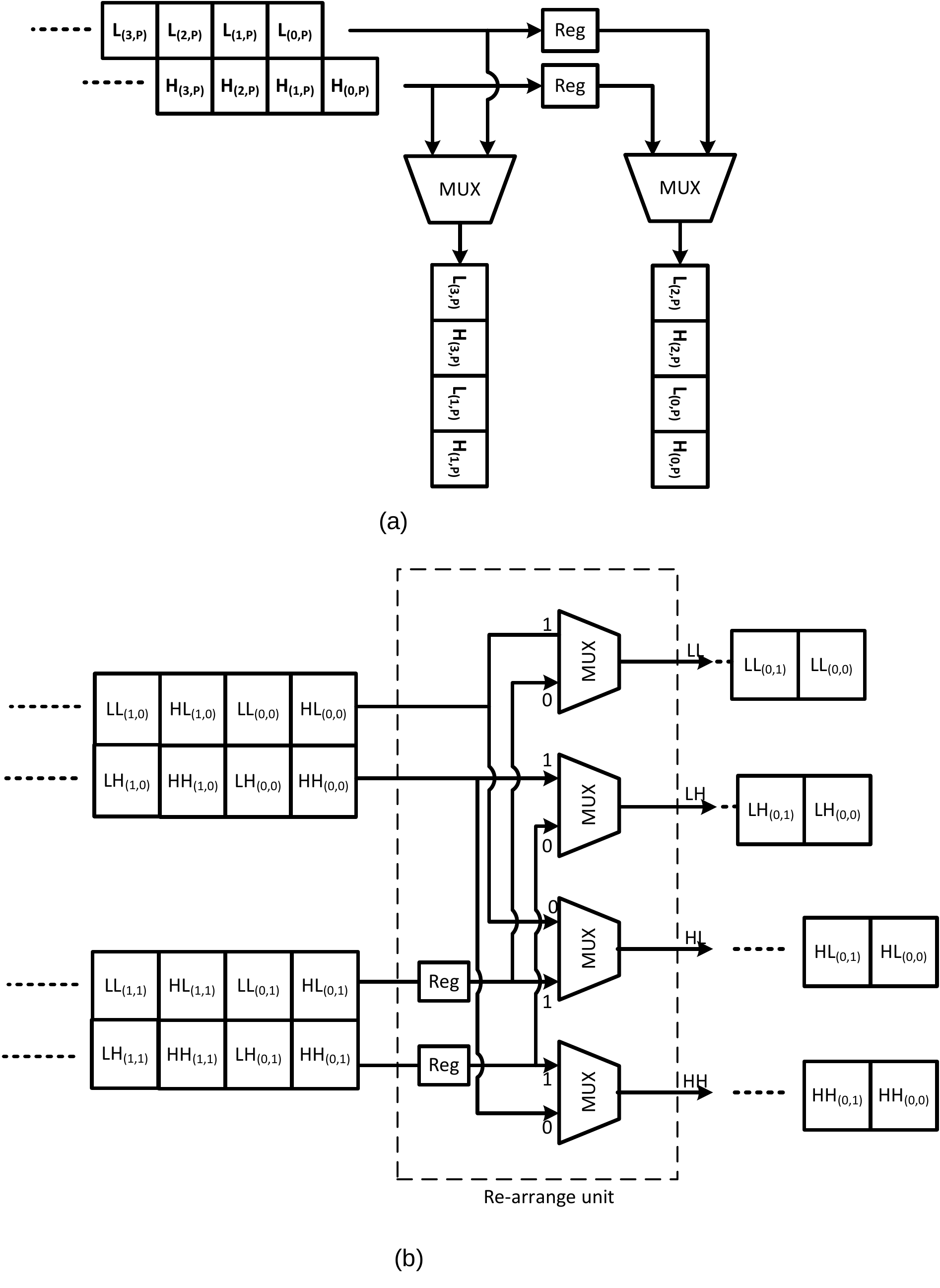}
\caption{(a) Transpose Register (Ref:\cite{3D_yusong2}) (b) Re-arrange Unit}
\label{3d_3}
\end{figure}

\subsubsection{Row Processor (RP)}
Let $ X $ be the image of size $N \times N$, extend this image by one column by using symmetric extension. Now image size is $N\times (N+1)$. Refer \cite{3D_yusong2} for the structure of strip based scanning method. The proposed architecture initiates the DWT process in row wise through row processor (RP) then process the column DWT by column processor (CP).  Fig.~\ref{3d_1}(a). shows the generalized structure for a row processor with $P $ number of PUs. $ P = 2 $ has been considered for our proposed design.  For the first clock cycle, RP get the pixels from $ X(0,0) $ to $ X(0,2P) $  simultaneously. For the second clock RP gets the pixels from next row i.e. $ X(1,0) $ to $X(1,2P)$, the same procedure continues for each clock till it reaches the bottom row i.e., $ X(N,0)$ to $ X(N,2P) $. Then it goes to the next strip and RP get the pixels from $ X(0,2P) $ to $ X(0,4P) $ and it continues this procedure for entire image. Each PU consists of five pipeline stages and each pipeline stage is processed by one processing element (PE) as depicted in Fig.~\ref{3d_2}(b). First stage (shift$\_$PE) provide the partial results which is required at $2^{nd}$ stage (PE$\_$alpha), likewise processing elements PE$\_$alpha to PE$\_$delta ($2^{nd}$ stage  to $5^{th}$ stage) gives the partial results along with their original outputs. (e.g., consider the PE$\_$alpha of PU-1, it needs to provide output corresponding to eqn.(1) ($ H_{1}[n]$), along with $H_{1}[n] $, it also provides the partial output $ X'[2n]$ which is required for the PE$\_$beta). Structure of the PEs are given in the Fig.~\ref{3d_2}(b), it shows that multiplication is replaced with the shift and add technique. The original multiplication factor and the value through the shift and add circuit are noted in Table.\ref{tab1}, it shows that variation between original and adopted one is extremely small. As shown in Fig.~\ref{3d_2}(b), time delay of shift$\_$PE is one $T_{a}$ and remaining all PEs are having delay of $ 2T_{a}$. To reduce the CPD of PU, PEs from PE$\_$alpha to PE$\_$delta are divided in to two pipeline stages, and each pipeline stage has a delay of $T_{a}$, as a result CPD of PU is reduced to $T_{a}$ and pipeline stages are increased to nine and is shown in Fig.~\ref{3d_2}(c). The outputs $ H_{1}[n+P-1]$, $ L_{1}[n+P-1]$, and $ H_{2}[n+P-1]$ corresponding to PE$\_$alpha and PE$\_$beta of last PU and PE$\_$gama of last PU is saved in the memories Memory$\_$alpha, Memory$\_$beta  and Memory$\_$gama respectively, shown in Fig.~\ref{3d_1}(a). Those stored outputs are inputted for next subsequent columns of the same row. For a $N\times N$ image rows is equivalent to $ N $. So the size of the each memory is $N\times 1$ words and total row  memory to store these outputs is equals to $ 3N $. Output of each PU are under gone through a process of scaling before it producing the outputs H and L. These outputs are fed to the transposing unit. The transpose unit has $ P $ number of transpose registers (one for each PU). Fig.~\ref{3d_3}(a) shows the structure of transpose register, and it gives the two H and two L data alternatively to the column processor. 

\subsubsection{Column Processor (CP)}

The structure of the column processor (CP) is shown in Fig.~\ref{3d_1}(b). To match with the throughput of RP, CP is also designed with  two number of PUs in our architecture. Each transpose register produces a pair of H and L in an alternative order and are fed to the inputs of one PU of the CP. The partial results produced are consumed by the next PE after two clock cycles. As such, shift registers of length two are needed within the CP between each pipeline stages for caching the partial results (except between $ 1^{st} $ and $2^{nd} $ pipeline stages). At the output of the CP, four sub-bands are generated in an interleaved pattern, $i.e. (HL,HH), (LL,LH), (HL,HH), (LL,LH),$ and so on. Outputs of the CP are fed to the re-arrange unit. Fig.~\ref{3d_3}(b) shows the architecture for re-arrange unit, and it provides the outputs in sub-band order $i.e.  LL, LH, HL$ and $HH$ simultaneously, by using $ P $ registers and $2P$ multiplexers. For multilevel decomposition, the same DWT core can be used in a folded architecture with an external frame buffer for the LL sub-band coefficients.
\begin{table}[]
\centering
\caption{Original and adopted values for multiplication}
\label{tab1}
\centering
\begin{tabular}{|l|l|l|}
\hline
     & Original  & Multiplier \\
PE       &  Multiplier &  value through  \\
     &  Value & shift and add \\
\hline
PE$ \_ $alpha & $ a'$=-0.6305      & $ a'$=-0.6328 \\
\hline
PE$ \_ $beta  & $ b'$=11.90        &$ b'$=12      \\
\hline
PE$ \_ $gama  &$ c'$=-21.378      & $ c'$=-21.375 \\
\hline
PE$ \_ $delta & $ d'$=2.55         & $ d'$=2.5625  \\
\hline
\end{tabular}
\end{table}   

\subsection{Architecture for Temporal Processor (TP)}

Eqn.(\ref{eq3}) shows that Haar wavelet transform depends on two adjacent pixels values (same pixel position of adjacent frames, for temporal processing). As soon as spatial processors are provide the 2-D DWT results, temporal processors starts processing on the spatial processor outputs (2-D DWT results) and produce the 3-D DWT results. Fig.~\ref{blockdia_1} shows that there is no requirement of temporal buffer, due to the sub-band coefficients of two spatial processors are directly connected to the four temporal processors. But it has been designed with 3 pipeline stages, it require 6 pipeline registers for each TP.  Same frequency sub-band of the distinct spatial processors are fed to the each temporal processor. i.e. $LL, HL, LH$ and $HH$ sub-bands of the spatial processor 1 and 2 are given as inputs to the temporal processor 1, 2, 3 and 4 respectively. Temporal processor apply 1-D Haar wavelet on sub-band coefficients, and provide the low frequency sub-band and high frequency sub-band as output. By combining all low frequency sub-bands and high frequency sub-bands of all temporal processors provide the 3-D DWT output in the form of L-Frame and H-Frame (2-D DWT by spatial processors and 1-D DWT by temporal processors).

\begin{table}
\centering
\caption{Device utilisation summary of the proposed architecture}
\label{FPGA_results}
\begin{tabular}{|l|c|c|c|}
\hline
Logic utilized       & Used                  & Available             & Utilization      \\ \hline
Slice Registers      & 1958                  & 106400                & 1\%                   \\ \hline
Number of Slice LUTs & 2852                  & 53200                 & 5\%                   \\ \hline
Number of fully      & \multirow{2}{*}{1137} & \multirow{2}{*}{3673} & \multirow{2}{*}{30\%} \\ 
used LUT-FF pairs    &                       &                       &                       \\ \hline
Number of Block RAM  & 3                     & 140                   & 2\%                   \\ \hline
\end{tabular}
\end{table}
\begin{table*}
\centering
\caption{Comparison of proposed 2-D DWT architecture with existing architectures (for 1-level)}
\label{2dcompare}
\begin{tabular}{|l|l|l|l|l|l|}
\hline
Parameter        & Zhang \cite{3D_zhang} & Mohanty \cite{3D_mohanty3}    & Darji \cite{2D_darji} & Yusong  \cite{3D_yusong2} & Proposed  \\ \hline
\hline
Multipliers      & 10    & 9P        & 10    & 10P      & 0         \\ \hline
Adders           & 16    & 16P       & 16    & 16P      & 34P       \\ \hline
Internal Memory  & 4N+37 & 15P+5.5N  & 4N    & 24P+3N   & 60P+3N \\ \hline
Critical path    & $T_{m}$    &$T_{m}+2T_{a}$   & $T_{m}$  & $T_{m}+T_{a}$    & $T_{a}$   \\ \hline
Computation Time & $ N^{2} $/2  & $ N^{2} $/2P     & $ N^{2} $/2  & $ N^{2} $/2P    & $ N^{2} $/2P     \\ \hline
Throughput       & 2/$T_{m}$ & 2P/$T_{m}+2T_{a}$ & 2/$T_{m}$  & 2P/$T_{m}+T_{a}$ & 2P/$T_{a}$    \\ \hline
\end{tabular}
\end{table*}
\begin{table*}
\centering
\caption{Comparison of proposed 3-D DWT architecture with existing architectures (for 1-level)}
\label{3dcompare}
\resizebox{\textwidth}{!}
{\begin{tabular}{|l|l|l|l|l|l|}
\hline
Parameters   & Weeks \cite{3D_weeks}        & Taghavi \cite{3D_tagavi} & A.Das  \cite{3D_swapna} & Darji \cite{3D_darji} & Proposed        \\ \hline
\hline
Memory requirement & $6N^{2} $+$ 6l$         & $5N^{2} $                   & $5N^{2} $   + 5N           & $4N^{2} $   + 10N        & 2*(3N+60P)+48              \\ \hline
Throughput/cycle   & -                               & 1 result                    & 2 results                  & 4 results                & 8 results       \\ \hline
Computing time & \multirow{2}{*}{$2N^{2}$ +  $ 3l$/2 }        & \multirow{2}{*}{$6N^{2} $} & \multirow{2}{*}{$3N^{2} $}             & \multirow{2}{*}{$3N^{2} $}           & \multirow{2}{*}{ $N^{2}$/2P } \\  For 2 Frames &            &                                  &           &        &             \\ \hline
Latency            & $2.5N^{2} $ + 0.5$ l $   & $4N^{2} $ cycles            & $2N^{2} $ cycles           & $3N^{2} $/2 cycles       & 21 cycles       \\ \hline
Area               & -                               & -                           & 1825 slices                & 2490 slices              & 2852 slice LUTs \\ \hline
Operating          & \multirow{2}{*}{200 MHz (ASIC)} & \multirow{2}{*}{-}          & 321 MHz                    & 91.87 MHz                & 265 MHz         \\ 
Frequency          &                                 &                             & (FPGA)                     & (FPGA)                   & (FPGA)          \\ \hline
Multipliers        & -                               & -                           & Nil                        & 30                       & Nil             \\ \hline
Adders             & $ 6l$ MACs                      & -                           & 78                         & 48                       &      168           \\ \hline
Filter bank        & $l$-length                      & D-9/7                       & D-9/7                      & D-9/7                    & D-9/7 (2-D) + Haar (1-D)           \\ \hline
\end{tabular}}
\end{table*}

\begin{table}
\centering
\caption{Synthesis Results (Design Vision) Comparison of Proposed 3-D DWT architecture with existing}
\label{3d_asic}
\begin{tabular}{|l|c|c|}
\hline
Parameters             & Darji et al.,\cite{3D_darji}     & Proposed    \\ \hline
Comb. Area     & 61351 $ \mu m^{2} $  & 526419 $ \mu m^{2} $  \\ \hline
Non Comb. Area & 807223 $ \mu m^{2} $ & 553078 $ \mu m^{2} $  \\ \hline
Total Cell Area        & 868574 $ \mu m^{2} $ & 1079498 $ \mu m^{2} $ \\ \hline
Operating Voltage      & 1.98 V     & 1.2 V        \\ \hline
Total Dynamic Power    & 179.75 mW  & 38.56 mW    \\ \hline
Cell Leakage Power     & 46.87 $ \mu W $   & 4.86 mW     \\ \hline
\end{tabular}
\end{table}

\section{Implementation Results and Performance Comparison}
 
The proposed 3-D DWT architecture has been described in Verilog HDL. A uniform word length of 14 bits has been maintained throughout the design. Simulation results have been verified by using Xilinx ISE simulator. We have simulated the Matlab model which is similar to the proposed 3-D DWT hardware architecture and verified the 3-D DWT coefficients. RTL simulation results have been found to exactly match the Matlab simulation results. The Verilog RTL code is synthesized using Xilinx ISE 14.2 tool and mapped to a Xilinx programmable device (FPGA) 7z020clg484 (zynq board) with speed grade of -3. Table \ref{FPGA_results} shows the device utilization summary of the proposed architecture and it operates with a maximum frequency of 265 MHz. 
\par The proposed architecture has also been synthesized using SYNOPSYS design compiler with 90-nm technology CMOS standard cell library. It consumes 43.42 mW power and  occupies an area equivalent to 231.45 K equivalent gate at frequency of 200 MHz.
\subsection{Comparison}
The performance comparison of the proposed 2-D and 3-D DWT architectures with other existing architectures is figure out in Tables \ref{2dcompare} and \ref{3dcompare} respectively. The proposed 2-D processor requires zero multipliers,  34P (Pis number of parallel PUs) adders, 60P+3N internal memory. It has a critical path delay of $T_{a}$ with a throughput of four outputs per cycle with $ N^{2} $/2P computation cycles to process an image with size $N \times N$.  When compared to recent 2-D DWT architecture developed by the Y.Hu et al. \cite{3D_yusong2}, CPD reduced to $T_{a}$ from $T_{m}+T_{a}$ with the cost of small increase in hardware resources. 
 
Table \ref{3dcompare} shows the comparison of proposed 3-D DWT architecture with existing 3-D DWT architecture. It is found that, the proposed design has less memory requirement, High throughput, less computation time and minimal latency compared to \cite{3D_weeks}, \cite{3D_tagavi}, \cite{3D_swapna}, and \cite{3D_darji}. Though the proposed 3-D DWT architecture has small disadvantage in area and frequency, when compared to \cite{3D_swapna}, the proposed one has a great advantage in remaining all aspects.   
 
 Table \ref{3d_asic} gives the comparison of synthesis results between the proposed 3-D DWT architecture and \cite{3D_darji}. It seems to be proposed one occupying more cell area, but it included total on chip memory also, where as in \cite{3D_darji} on chip memory is not included. Power consumption of the proposed 3-D architecture is very less compared to \cite{3D_darji}.
\section{Conclusions}
In this paper, we have proposed memory efficient and high throughput architecture for lifting based 3-D DWT. The proposed architecture is implemented on 7z020clg484 FPGA target of zynq family, also synthesized on Synopsys' design vision for ASIC implementation. An efficient design of 2-D spatial processor and 1-D temporal processor reduces the internal memory, latency, CPD and complexity of a control unit, and increases the throughput. When compared with the existing architectures the proposed scheme shows higher performance at the cost of slight increase in area. The proposed 3-D DWT architecture is capable of computing 60 UHD (3840$ \times $2160) frames in a second.

\end{document}